\documentclass[fleqn,usenatbib]{mnras}

\usepackage{newtxtext,newtxmath}

\usepackage[T1]{fontenc}
\usepackage{ae,aecompl}
\usepackage{booktabs}
\usepackage{multirow}
\usepackage{siunitx}
\usepackage{rotating}

\usepackage{graphicx}	
\usepackage{amsmath}	
\usepackage{amssymb}	
\usepackage{float}

\def \kpc         {\rm kpc}

\def \msun        {M_\odot}

\title[Giant LSB galaxy probes MOdified Gravity]{Giant low-surface-brightness dwarf galaxy as a test bench for MOdified Gravity}

\author[I. de Martino et al]{
Ivan de Martino$^{1,2,3}$\thanks{E-mail: ivan.demartino1983@gmail.com}
\\
$^{1}$ Dipartimento di Fisica, Universit\`a di Torino,  Via P. Giuria 1, I-10125 Torino, Italy\\
$^{2}$ Istituto Nazionale di Fisica Nucleare (INFN), Sezione di Torino, Via P. Giuria 1, I-10125 Torino, Italy\\
$^{3}$ Donostia International Physics Center (DIPC), 20018 Donostia-San Sebastian (Gipuzkoa) Spain
}

\date{Accepted XXX. Received YYY; in original form ZZZ}

\pubyear{2018}

\begin{document}
\label{firstpage}
\pagerange{\pageref{firstpage}--\pageref{lastpage}}
\maketitle

\newcommand{\aea}{Astron. Astrophys.}
\newcommand{\cqg}{Class.  Quant. Grav.}
\newcommand{\grg}{Gen.  Rel. Grav.}
\newcommand{\plb}{Phys. Lett. B}


\begin{abstract}
The lack of detection of supersymmetric particles is leading to look at alternative avenues for explaining  dark matter's effects. Among them, modified theories of gravity may play an important role accounting even for both dark components needed in the standard cosmological model. Scalar-Tensor-Vector Gravity theory has been proposed to resolve the dark matter puzzle. Such a modified gravity model introduces, in its weak field limit, a Yukawa-like correction to the Newtonian potential, and  is capable to explain most of the phenomenology related to dark matter at scale of galaxies and galaxy clusters. Nevertheless, some inconsistencies appears when studying systems that are supposed to be dark matter dominated such as dwarf galaxies. In this sense Antlia II, an extremely diffuse galaxy which has been recently discovered in {\em Gaia}'s second data release, may serve  to probe the aforementioned theory against the need for invoking dark matter. Our analysis shows several inconsistencies and leads to argue that MOdified Gravity may not be able to shed light on the intriguing  nature of dark matter.
\end{abstract}
\begin{keywords}
dark matter -- galaxy, kinematics and dynamics -- galaxies, dwarf-- gravitation -- data analysis
\end{keywords}

\section{Introduction}
Nowadays, General Relativity has been widely verified by a plethora of observations (\citet{ Stairs2003, Everitt2011,will_2018}). Nevertheless,  datasets with unprecedented accuracy have been exploited to unveil the mysteries behind the need to invoke dark matter and dark energy to explain the dynamics of self-gravitating systems, the gravitational lensing and the accelerated expansion of the Universe among the others \cite{Blake2011,Hinshaw2013,Planck16_13}. 
Dark matter is understood to be an heavy particle beyond standard particle physics \citep{Bertone2018}.  However, the lack of direct detection of such a particle (\citet{lux2017,Xenon1T}) is increasingly favouring alternative candidates,  like QCD axions and ultra-light axions (\citet{Schive2014,Capolupo2016,demartino2017b,Capolupo2018,capolupo2019}), having very light masses, and being able to explain the small scale features of  dwarf galaxies  \citep{Moore1994}.
Albeit dark matter is still favoured over more exotic alternative, 
one should also mention that there is another viable way to solve the puzzle. In the last decades, many extensions/modifications of General Relativity have also been proposed to replace the need for dark energy with a curvature effect arising from the higher order terms and/or scalar fields that are introduced in those theories (\citet{mof1, PhysRept}). Besides, some theories may also account for dark matter. In fact, the Newtonian gravitational potential 
turns out to be modified opening to the possibility of explaining galaxy rotation curve, gravitational lensing and galaxy clusters without accounting for dark matter (\citet{Nojiri2011,  mof2, mof3, idm2015,manos, mof9}).

Among these theories, Scalar-Tensor-Vector Gravity theory (STVG), or more briefly MOdified Gravity (MOG) theory, has been proposed to specifically resolve the dark matter puzzle (\citet{mof1,mof2,mof3}). Such a theory of gravity extends the standard Hilbert-Einstein action adding adds scalar, tensor and vector fields. Since the matter is coupled to such massive vector fields, test particles do not follow geodesics, and a repulsive gravitational force at short range ({\em i.e.} galactic, sub-galactic, or smaller scales) arises reducing the total strength of the gravitational interaction. Recently, MOG has been tested at both galactic and extra-galactic scales showing the capabilities to fit mass-radius relation of compact stars, galactic rotation curves, multi-wavelengths observations of galaxy clusters, cosmological datasets, and the direct detection of gravitational waves
(\citet{mof4,mof6,mof2,mof3,mof7,mof5,mof8,mof9}). Interestingly, one should also note that \citet{mof10} carried out an analysis of dynamics of stars in dwarf spheroidal galaxies pointing out several inconsistencies  that needs to be verified. Specifically, although they were analyzing galaxies in the same range of masses and luminosities, \citet{mof10} did not found a common value for two free parameters ascribed to the gravitational theory. In other words, they show the need to fit the model case-by-case which led to argue that the gravitational model in not able to self-consistently fit the dynamics of  dwarf galaxies. 

Recently, using the second data release of the astrometric satellite {\em Gaia},  it has been discovered a very diffuse, cored, and low surface brightness dwarf galaxy in orbit around Andromeda \citep{Torrealba2019}. Antlia II has a core radius of  $\sim 2.9$ kpc and an exceptionally small velocity dispersion $\simeq 6$ km/s. Its low measured surface brightness ($ 32.3 mag/arcsec^2$) makes it $\sim100$ times more diffuse than the ultra diffuse galaxies. Since the dynamics of stars in these extremely low-surface-brightness dwarfs is  dominated by the dark matter distribution, it has been argued that the formation of such a large core may  require alternatives to the standard cold dark matter, such as ultra-light bosons \citep{idm2019}. Thus, Antlia II represents an exceptional laboratory where to carry out tests of modified gravity and, in particular, of those theories that aim to avoid the introduction of a dark matter component, such as MOG. In order to explore its capability to explain these type of galaxies avoiding the need of a dominant dark matter core, we predict the dispersion velocity profile in the framework of MOG theory and fits it to the data from Antlia II. 

The next section will be devoted to explain hot to predict the dispersion velocity profile in MOG theory. Then, we will show the results of our analysis and give our conclusions.

\section{MOdified Gravity  and Stellar Dynamics}\label{sect:data}

The MOG Lagrangian accounts for the general relativistic term including the cosmological constant, plus additional terms for the scalar and vector fields \citep{mof1}. From such a generalization of the Einstein-Hilbert Lagrangian,  
one obtains the following field equations \cite{mof1,mof2,mof3}:
\begin{equation}
G_{\mu\nu}+G\left(\square \frac{g_{\mu\nu}}{G}-\nabla_\mu \nabla_\nu \frac{1}{G} \right)=8\pi G T_{\mu\nu}^{(tot)}\,,
\label{1a}
\end{equation}
where  $ T_{\mu\nu}^{(tot)} = T_{\mu\nu}^{(m)} + T_{\mu\nu}^{(\phi)} + T_{\mu\nu}^{(s)}$ is the total energy-momentum tensor, and $T_{\mu\nu}^{(m)}$, $T_{\mu\nu}^{(\phi)}$ and $T_{\alpha\beta}^{(s)}$ are the energy momentum tensors for ordinary matter, vector and scalar fields, respectively. Furthermore, G is space-time dependent scalar field, usually defined as $G(x)=G_N(1+\alpha(x))$, where $G_N$ is the Newtonian gravitational constant, and $\square=g^{\mu\nu}\nabla_a\nabla_b$ is the D'Alembertian operator.

In order to compute the weak filed limit, and to obtain a modified version of the Poisson's equation,  the field equations in Eq. \eqref{1a} must be linearised. This leads to an analytic expression for the gravitational potential (\citet{mof1,mof2, mof3})
\begin{align}
	\Phi\left(\textbf{r}\right)=&-G_N\left(1+\alpha\right)\int\frac{\rho\left(\textbf{r}'\right)}{\mid\textbf{r}-\textbf{r}'\mid}d^3r'\nonumber\\
	&+G_N\alpha\int\frac{\rho\left(\textbf{r}'\right)}{\mid\textbf{r}-\textbf{r}'\mid}e^{-\mu\mid\textbf{r}-\textbf{r}'\mid}d^3r'\,,
	\label{1b}
\end{align}
where $\mu$ is the inverse of the characteristic length of the gravitational potential, $\alpha\equiv\frac{G_{\infty}-G_N}{G_N}$, and $G_\infty$ is the  effective gravitational constant at infinity, respectively. Let use note that  the Newtonian potential is recovered by setting $\alpha=0$.

There are only two free parameters in MOG theory, namely $\alpha$ and $\mu$. Thus, previous analysis based on the study of galactic rotation curves found tight constraints: $\alpha = 8.89\pm0.34$ and $\mu=0.042\pm0.004 \kpc^{-1}$, respectively \citep{mof2}. Also, it was argued that, with those values, MOG theory would have been able to describe the dynamics of all self-gravitating systems. So that, they were declared to be {\em universal} constants. Nevertheless, those parameters slightly depend by the mass of the systems. As a consequence, their values needed to account for the Sunyaev-Zeldovich emission in galaxy clusters differs from the {\em universal} ones \citep{mof9}. Furthermore, \citet{mof10} analyzed the dispersion velocity profile of several dwarf spheroidals without finding a common set of parameters to describe them, and pointing out a strong tension between the best fit value of  $\alpha$ and $\mu$ in each dwarfs and the {\em universal} values of those parameters.  
Since the  two parameters $\alpha$ and $\mu$ are expected to depend by the total mass of the system,  they could also differ to fit the dispersion velocity in Antlia II, whose estimated total mass is at least one order of magnitude less than the one of the dwarf galaxies studied in \citet{mof10}.

Using Eq. \eqref{1b}, it is straightforward to compute the dispersion velocity of stars within the modified gravitational potential well by solving the spherical Jeans equation:
\begin{equation}
\frac{d(\rho_*(r)\sigma_r^2(r))}{dr} = -\rho_*(r)\frac{d\Phi(r)}{dr}-2\beta\frac{\rho_*(r)\sigma_r^2(r)}{r},
\end{equation}
where $\rho_*(r)$ is the tracer star's density, and $\beta$ is the anisotropy parameter(\cite{BT}, Equation (4.61)).
The general solution of the Jeans's equation is given by
\begin{equation}
\rho_*(r)\sigma^2_r(r)=r^{-2\beta}\int_{r}^{\infty} \frac{d\Phi(x)}{dx}\rho_*(x)x^{2\beta} \,dx \,.
\end{equation}

{ Here, following \citep{Torrealba2019}, we model the star density as Plummer distribution}
\begin{equation}
\rho_*(r) =  \frac{\rho_{p,0}}{\left(1+\frac{r^2}{r_p^2}\right)^{5/2}}\,,
\label{Plummer}
\end{equation}
where $\rho_{p,0}$ is computed so that the total stellar mass at the half-light radius $r_h$  has to match the observed one, $M(r<r_h) =(5.4\pm2.1)10^7\msun$  \citep{Torrealba2019}. Usually, the scale radius ($r_p$) is assumed to coincide with the observed half-light radius $r_h$.

Finally, to directly compare the dispersion velocity with the data, we project the velocity dispersion along the line of sight as follows
\begin{equation}
\sigma^2_{los} (R) = \frac{2}{\Sigma(R)}\int_{R}^{\infty}\biggl(1-\beta\frac{R^2}{r^2}\biggr) \frac{\sigma^2_r(r)\rho_*(r)}{(r^2-R^2)^{1/2}} r \,dr\,,
\end{equation}
where R is the projected line-of-sight radius, and $\Sigma(R)$ is the surface brightness of the galaxy given by
\begin{equation}
\Sigma(R) = 2\int_{R}^{\infty} \frac{\rho_*(r)}{(r^2-R^2)^{1/2}} r dr\,.
\end{equation}

We should note that, previous analysis based on the Navarro-Frenk-White dark matter density profile \citep{Torrealba2019}, and on the wave-like dark matter model \citep{idm2019}, strongly favours $\beta=0$ (isotropy). Nevertheless, we will relax this hypothesis in our model to understand the impact of the anisotropic parameter. 

\section{Results}

The dispersion velocity profile of Antlia II dwarf galaxy has been recently measured in the second data release of  the satellite {\em Gaia} \citep{Torrealba2019}. We have solved the spherical Jeans equation using the modified Newtonian potential in Eq \eqref{1b}. Then, the result has been projected along the line of sight to be compared with the data. We carried out two independent analysis: in model (A) we set $r_p=r_h$ retaining the observed values of the half-light radius $r_h=2867\pm312$ pc \citep{Torrealba2019}, while we varied both parameters of the MOG theory, $(\alpha, \mu)$, and the anisotropic parameter $\beta$; in model (B), instead, we assumed isotropy of the system ($\beta=0$), and vary the value of the scale radius of the Plummer profile together with   $(\alpha, \mu)$. The parameter space was explored using a Monte Carlo Markov Chain (MCMC)  pipeline based on the Metropolis-Hastings sampling  algorithm. The step size is adapted to reach an optimal acceptance rate between 20\% and 50\%, while the convergence has been guaranteed employing the Gelman-Rubin criteria (\citet{Metropolis1953,Hastings1970,Gelman1992,Gelman1996,Roberts1997}). We used flat priors covering the range $[0.0, 30] $ for $\alpha$, $[0.0, 0.7]$ for the inverse scale length $\mu$ (which has unit of $kpc^{-1}$) and, finally, $[1.0, -100] $ for the anisotropic parameter $\beta$ in the model (A), and 
 $[0.0, 10]$ kpc for the scale radius of the Plummer profile $r_p$ in the model (B). Then,  we run four independent chains with random starting points,  containing a total number of at least $80,000$ steps, and stopping when the convergence is reached. Afterwards, we merged together the different chains to compute the expectation value and  variance for each parameter.  

From our analysis, we have obtained the following bounds on the parameters. In the case of the Model (A), we find  $\alpha=3.28^{+1.89}_{-1.73} $,  $\mu(10^{-2} \kpc^{-1})=14.21^{+10.15}_{-9.61} $, and  $\beta=-18.22^{+12.44}_{-12.08} $, respectively. While in the case of the Model (B), we get  $\alpha=6.16^{+9.00}_{-3.66} $,  $\mu(10^{-2} \kpc^{-1})=18.78^{+8.16}_{-10.72} $, and  $r_p (\kpc)=1.68^{+0.71}_{-0.18} $, respectively.
The best fit values and the errors at the 68\% of the confidence level have also been summarized in Table 1.
\begin{table}
	\begin{tabular}{|c|c|c|c|c|}\hline
	    & $\alpha$ & $\mu$ & $\beta$ & $r_p$\\[0.1cm]
		&          & $(10^{-2} \kpc^{-1})$ & &(\kpc)\\[0.1cm]
		\cline{1-5}
		& & & &\\[-0.2cm]
		Model A & $3.28^{+1.89}_{-1.73}$ & $14.21^{+10.15}_{-9.61}$ & $-18.22^{+12.44}_{-12.08}$ &- \\[0.1cm]
		\cline{1-5}
		& & & &\\[-0.2cm]
		Model B
		& $6.16^{+9.00}_{-3.66}$ & $18.78^{+8.16}_{-10.72}$ & -& $1.68^{+0.71}_{-0.18}$ \\[0.1cm]
		\hline
	\end{tabular}
\caption{The table summarizes the best fit values of the parameters, and their 68\% uncertainty for two models. In the Model A we fixed the value of the Plummer scale $r_p$ to the measured value of the half-light radius while fitting the anisotropic parameter $\beta$. In the Model B, we fixed $\beta=0$ while varied $r_p$.}
\end{table}

\begin{figure}
\centering
\includegraphics[width=0.95\columnwidth]{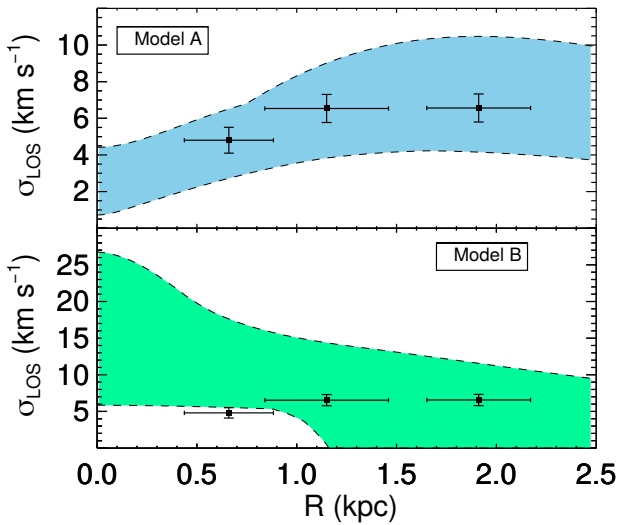}
\caption{In the figure we have depicted the results of our analysis for Model A and B, in the upper and lower panel, respectively. The coloured strips show the region within the 68\% of the confidence level.}\label{fig1}
\end{figure}

In Figure \ref{fig1}, we have illustrated the effectiveness of MOG theory in fitting the dispersion velocity profile of Antlia II. The coloured strips represent the regions of the parameter space within the 68\% of the confidence level. In the case of the Model A, we have fixed the Plummer scale to the measured half-light radius, as it is in \citet{Torrealba2019, idm2019}, while we left $\beta$ free to vary. In such a way, as shown in the upper panel of Figure \ref{fig1}, we are able to fit the dispersion velocity profile emulating, in the framework of MOG, the extended dark matter core needed in standard gravity \citep{Torrealba2019, idm2019}. However, the results need a deeper interpretation. First, our best fit value of the two MOG's parameters differs from the {\em universal} values found in \citet{mof2} at more then $3$ and $1.5$ $\sigma$ for $\alpha$ and $\mu$, respectively. Since the {\em universal} values were obtained by fitting the galactic rotation curves of more massive galaxies, this small tension was rather expected. Nevertheless, the tension in $\alpha$ increases when considering the best fit value in dwarf spheroidals. Specifically,  \citet{mof10} found a value of $\alpha=211.4^{+46.1}_{42.7}$ for Fornax (representing their best case in term of reduced-$\chi^2$) which points out a tension to a level of $4.5\sigma$, indicating strong difficulties of MOG to self-consistently account for the dynamics of stars in dark matter dominated systems such as dwarf galaxies. Second, the best fit value of the anisotropic parameter differs from zero (isotropy) at more than $1.5\sigma$. This results is rather controversial. It means that to fit the dispersion velocity profile we need the tangential dispersion velocity to be much higher than the radial velocity, against the hypothesis made in \citet{Torrealba2019,idm2019}, and tested in our Model (B).  Also, having such a tangential bias would requires that this dwarf galaxy was populated by mainly stars in quasi-circular orbits, losing ergodicity and, hence, lowering the dispersion velocity profile toward the center \citep{BT}, against the central observed flatness of the dispersion velocity. { Such a configuration is strongly disfavoured in a dwarf galaxy. Nevertheless, the error on the anisotropy parameter is $\sim 66\%$ which leaves a window for a much lower best fit value which may reconcile the model with the observational dataset and the dwarf nature of the galaxy.}

Considering the aforementioned difficulties of MOG, we have also carried out an additional analysis.  We set $\beta=0$, forcing the model to be isotropic, but allowing the Plummer scale to vary. First, the tension of $\alpha$ and $\mu$ with their {\em universal} values is reduced to $1\sigma$ level in both parameters, while the huge tension with other independent analysis of dwarfs is confirmed. Second, the value of the Plummer scale is at $1.5\sigma$  from the observed half-light radius and, hence,  leads to slightly different value of the enclosed mass: $M(<r_h)=3.17^{+1.82}_{-1.27}\times10^7M_\odot$, which is anyway compatible at $1\sigma$ with the estimation of \citet{Torrealba2019}. Finally, and most importantly, the model is not able to reproduce the dispersion velocity profile. As shown in the bottom panel of Figure \ref{fig1}, under those assumptions, the model never emulate the effect of a dark matter core. If the dispersion velocity profile get flatter below $0.5\kpc$, then it decrease quickly at outermost radii. On the contrary, when the outer data points are fitted, the dispersion velocity is more cuspy in the innermost region till reaching $\sim 25 km/s$, which is a factor 5 times higher than the measured values. 

The aforementioned considerations lead to argue that MOG theory fails in reproducing the dynamics of stars in Antlia II without introducing in the orbital distribution a strong tangential bias. This results strongly support the previous ones showed in \citet{mof10}, where several inconsistencies in describing the dynamics of stars in dwarf spheroidals were already highlighted. 

\section{Discussions and Conclusions}

The lack of detection of supersymmetric particles at LHC led alternative candidates for dark matter to rapidly gaining attention. Apart from other particles, modifications of the gravitational action also offer a new avenue to account for the dark matter. Among many theories, MOG has been successfully tested with galaxies and cluster of galaxies showing the capability to explain the rotation curves of spiral galaxies, the gravitational lensing, the Sunyaev-Zeldovich effect and the X-ray emission in galaxy clusters without resorting to any dark matter component (\citet{mof4,mof6,mof2,mof3,mof7,mof5,mof8,mof9}). Nevertheless,  a recent study of the dynamics of dwarf spheroidals orbiting around the Milky Way has shown some inconsistencies with previous results \citep{mof10}. Since dwarfs are supposed to be dominated by dark matter, any theory that aims to replace it with a modification of the gravitational potential must also be able to account for the dynamics of these galaxies. 
Here, we have investigated the dynamics of stars in Antlia II in the framework of MOG theory. Antlia II is a recently discovered low-surface-brightness dwarf  galaxy being $\sim 100$ more diffuse than ultra diffuse galaxies. It has a very wide core $\sim 3$ kpc, and it is supposed to be strongly dominated by dark matter (\citet{Torrealba2019, idm2019}). Due to its own nature, it represents an ideal candidate to test alternative theories of gravity such as MOG. Therefore, we have predicted the dispersion velocity profile by solving the spherically symmetric Jeans Equation, and assuming that the dynamics of stars is determined by the modified potential well in Eq. \eqref{1b} arising in MOG weak field limit, and the stellar density profile  is well described by the  Plummer model in Eq. \eqref{Plummer}. Finally, we have projected the solution of Jeans Equation along the line of sight to compare it with the data. 

We have carried out two analysis. In the first one, renamed Model A, we varied the parameters of MOG theory, $\alpha$ and $\mu$, plus the anisotropic parameter, $\beta$, while we set the scale length of the Plummer profile to the measured half-light radius. In the second analysis, labeled Model B, we set $\beta=0$ allowing the scale radius of the stellar profile to vary. In both case we found a discrepancy between the best fit values of the MOG parameters we obtained and the one obtained in other independent analysis. 
Of greater importance is that, Model A needs a strong tangential bias in the orbital distribution to reproduce  the dispersion velocity profile, while Model B is unable to reproduce it in the innermost region ($r<0.5\kpc$). These inconsistencies strongly support the results in \citet{mof10}, and led us to argue that MOG is unable to reproduce the dynamics of stars in Antlia II, as described in the previous section. Nevertheless, more data are needed to definitively  
ruled out the theory. Ideally, forthcoming data-sets collected by the next generation surveys such as Maunakea Spectroscopic Explorer (MSE, \citet{Bechtol2019} and Large Synoptic Survey Telescope (LSST, \citet{lsst2019}), will achieve the sensitivity needed to discover many more galaxies with similar features, and will allow   a detailed study of alternative theories of gravity in such dark matter dominated systems. Besides, future astrometric satellite primarily designed to study the local dark matter properties of the faintest objects in the Universe \citep{theia} will measure the proper motion of stars allowing to definitively rule out the model.

\section*{Acknowledgement}
IDM is supported by the grant ”The Milky Way and Dwarf Weights with Space Scales” funded by University of Torino and Compagnia di S. Paolo (UniTO-CSP). IDM also acknowledges partial support from the INFN grant InDark and the Italian Ministry of Education, University and Research (MIUR) under the Departments of Excellence grant L.232/2016

\bibliographystyle{mnras}
\bibliography{dwarf_mog}

\label{lastpage}
\end{document}